\documentclass{svmult}
\usepackage{default}
\usepackage{graphicx}
\usepackage{amssymb}
\usepackage{amsmath}
\def\be{\begin{equation}}
\def\ee{\end{equation}}

\def\deg{{\rm deg}}
\newtheorem{thrm}{\bf Theorem}
\newtheorem{lmm}{\bf Lemma}
\newtheorem{rmk}{\bf Remark}
\newtheorem{dfn}{\bf Definition}

\begin{document}

\title*{Recent Progress in Optimization of\\ Multiband Electrical Filters}
\author{Andrei Bogatyr\"ev
}
\institute{Andrei Bogatyr\"ev \at Institute for Numerical Math., Russian Academy of Sciences, 
Moscow ul. Gubkina, 8 Russia 119333 \email{ab.bogatyrev@gmail.com}}
\titlerunning{Multiband filters}
\authorrunning{Andrei Bogatyr\"ev}
\date{}
\maketitle

\abstract{The best uniform rational approximation of the \emph{sign} function on two
intervals was explicitly found by Russian mathematician E.I. Zolotar\"ev
 in 1877.  The progress in math eventually led to the progress in technology:
half a century later German electrical engineer and  physicist
W.Cauer has invented low- and high-pass electrical filters known today as
elliptic or Cauer-Zolotar\"ev filters and possessing the unbeatable quality. We discuss a recently developed
approach for the solution of optimization problem naturally arising in the synthesis of
multi-band  (analogue, digital or microwave)  electrical filters.  The
approach is based on techniques from algebraic geometry and
generalizes the effective representation of Zolotar\"ev fraction.}

\section{History and background}
Sometimes the progress in mathematics brings us to the progress in technology.
One of such examples is the invention of low- and high- pass electrical filters widely used  
nowadays is electronic appliances. The story started in year 1877 when E.I. Zolotar\"ev (1847-1878)  -- the pupil of P.L.Chebysh\"ev -- 
has solved a problem of best uniform  rational approximation of the function $sign(x)$ on two segments of 
real axis separated by zero. His solution now called \emph{Zolotar\"ev fraction} is the analogy of Chebysh\"ev polynomials in the realm of rational functions and inherits many nice properties of the latter. This work of Zolotar\"ev who also attended lectures of K.Weierstrass and 
corresponded to him was highly appreciated by the German scholar. More than 50 years later German electrical engineer, physicist and 
guru of network synthesis Wilhelm Cauer (1900-1945) has invented electrical filters with the transfer function
based on Zolotar\"ev fraction.

Further development of technologies brings us to more sophisticated optimization problems \cite{CWC,2,7,8,11}. In particular, 
modern gadgets may use several standards of wireless communication like IEEE 806.16, GSM, LTE, GPS..
and therefore a problem of filtering on several frequency bands arises. Roughly, the problem is this:
given the mask of a filter, that is the boundaries of its stop and pass bands, the levels of attenuation at 
the stopbands and the permissible ripple magnitude at the passbands, to find minimum degree real rational function fitting this mask.
The problem reduces to a solution of a series of somewhat more simple minimal deviation problems on several segments similar to the one considered by Zolotar\"ev. Several equivalent formulations will appear in Sect. \ref{4Set}

Those problems turned out to be very difficult from the practical viewpoint because of intrinsic instability of most numerical methods of rational approximation. However, we know how the 'certificate' of the solution (see contribution from Panos Pardalos in this volume) for this particular case looks like: the solution possesses the so called \emph{equiripple property}, 
that is behaves like a wave of constant amplitude on each stop or passband. The total number of ripples is bounded from below. In a sense the solution for this problem is rather simple -- you just manifest function with a suitable equiripple property.
Such behaviour is very unusual for generic rational functions therefore functions with equiripple property fill in a variety 
of relatively small dimension in the set of rational functions of bounded degree. The natural idea is to look for the solution in the 'small' set of the distinguished functions instead of the 'large' set of generic functions.
\emph{Ansatz} is an explicit formula with few parameters which allows to parametrize the 'small' set. This Ansatz ideology 
had been already used to calculate the so called Chebysh\"ev polynomials on several segments \cite{B99}, optimal stability polynomials for explicit multistage \emph{Runge-Kutta} methods \cite{B04,B05} and solve some other problems.  Recall e.g. \emph{Bethe Ansatz} for finding exact solutions for  Heisenberg antiferromagnetic.  Ansatz for optimal electrical filters is discussed in Sect.  \ref{Ansatz}.

\section{Optimization problem for multiband filter}
Suppose we have a finite collection $E$ of disjoint closed segments of real axis $\mathbb{R}$.
The set has a meaning of frequency bands and is decomposed into two subsets: $E=E^+\cup E^-$
which are respectively called the passbands $E^+$ and the stopbands $E^-$. Both subsets are non empty.
Optimization problem for electrical filter has   several equivalent settings \cite{Cauer, AS, Akh, Zol, Malo}

\subsection{Four settings}\label{4Set}
In each of listed below cases we minimize certain quantity among real rational functions $R(x)$ of 
bounded degree $\deg R\le n$ being the maximum of the  degrees of numerator and denominator of the fraction. 
The goal function may be the following

\subsubsection{Minimal deviation}\label{minDev}
$$\frac{\max_{x\in E^+}|R(x)|}{\min_{x\in E^-}|R(x)|}\longrightarrow \min =:\theta^2\le1$$

\subsubsection{Minimal modified deviation}\label{minDev2}
$$\max\{\max_{x\in E^+}|R(x)|,\max_{x\in E^-}1/|R(x)|\}\longrightarrow \min =:\theta$$

\subsubsection{Third Zolotar\"ev problem}\label{Zolo3} 
Minimize $\theta$ under the condition that there exist real rational function $R(x)$, 
$~\deg R\le n$, with the restrictions
	$$\min_{x\in E_-}|R(x)|\ge\theta^{-1},\quad \max_{x\in E_+}|R(x)|\le\theta$$

\subsubsection{Fourth Zolotar\"ev problem} \label{Zolo4}
Define the indicator function $S(x)=\pm1$ when $x\in E^\pm$. Find the best uniform rational
approximation $R(x)$  of $S(x)$ of the given degree: 
  $$ 
 ||R-S||_{C(E)} := \max\limits_{x \in E}|R(x) - S(x)| \to \min =: \mu.
  $$	
  
It is a good exercise to show that all four settings are equivalent and in particular the value of $\theta$ is the same for the first three settings and $1/\mu=(\theta+1/\theta)/2$ for the fourth one.

\subsection{Study of optimization problem}
Setting \ref{minDev} appears in the paper \cite{AS} by R.A.-R.Amer, H.R.Schwarz (1964). It was transformed to 
\ref{minDev2} by V.N.Malozemov (1979) \cite{Malo}. Setting \ref{Zolo3} appears after suitable normalization of the rational function 
in \ref{minDev} and essentially coincides with the third Zolotar\"ev problem \cite{Zol}. Setting \ref{Zolo4} coincides with the fourth Zolotar\"ev problem \cite{Zol} and was studied by N.I.Akhiezer \cite{Akh} (1929). The latter noticed that already in the classical Zolotar\"ev case when the set $E$ contains just two components,  the minimizing function is not unique. This phenomenon was fully explained in the dissertation of R.-A.R.Amer \cite{AS} who decomposed the space of rational functions of bounded deviation (defined in the left hand side of formula) in \ref{minDev} into classes. Namely, it is possible to fix the sign of the polynomial in the numerator of the fraction on each stopband and fix the sign of denominator polynomial on each passband. Then in the closure of each nonempty class there is a unique minimum. All mentioned authors established  that (local) minimizing functions are characterized by \emph{alternation} (or \emph{equiripple} in terms of electrical engineers) property. For instance, in the fourth Zolotar\"ev problem the approximation error $\delta(x):=R(x)-S(x)$ of degree $n$ minimizer has $2n+2$ \emph{alternation} points $a_s\in E$ where  $\delta(a_s)=\pm||\delta||_{C(E)}$ with consecutive change of sign.

\section{Zolotar\"ev fraction}
\begin{figure}
\centerline{\includegraphics[scale=.1]{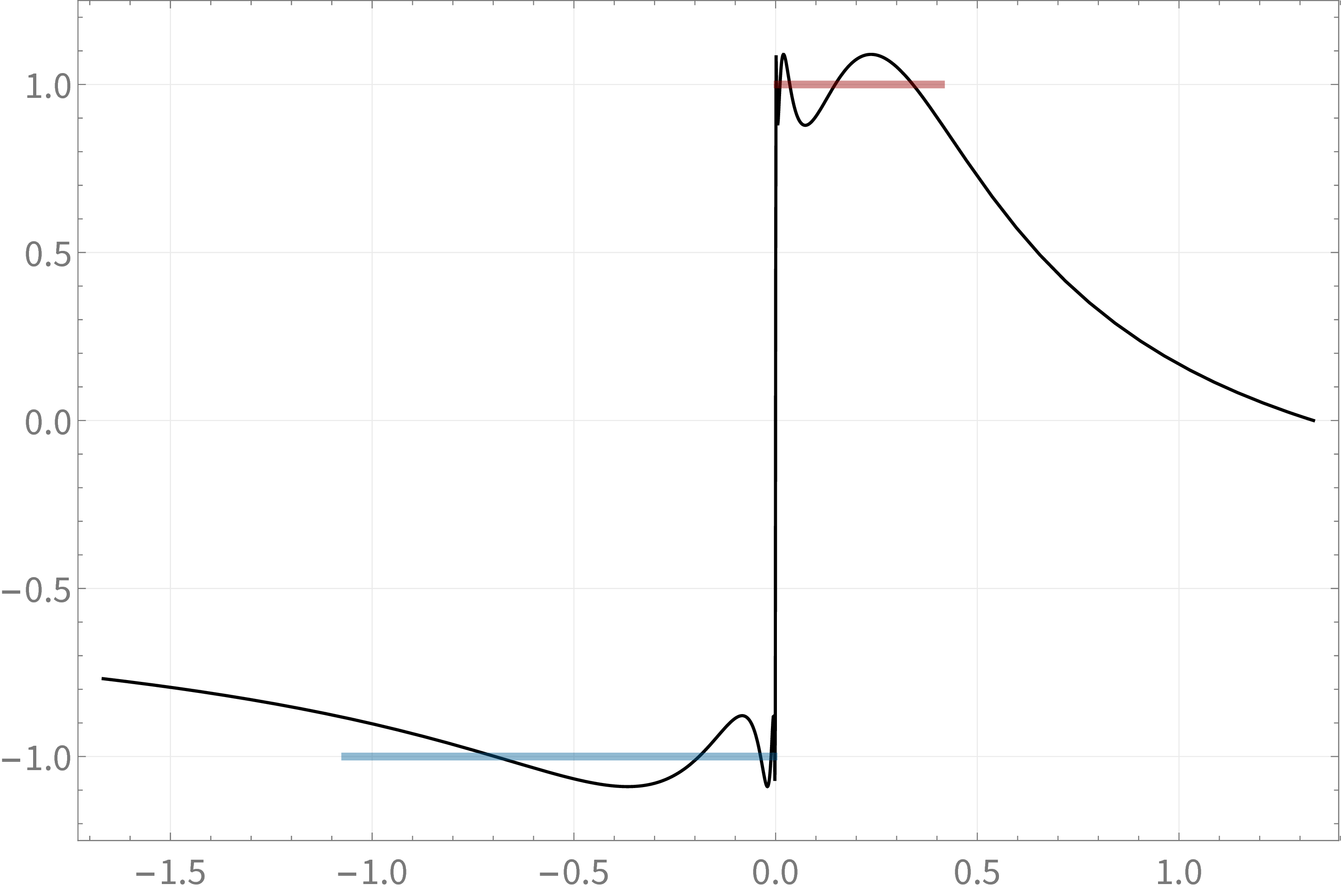}}
\caption{Graph of Zolotar\"ev fraction adapted to two segments $E^\pm$ of different lengths}
\label{ZF}
\end{figure}

E.I.Zolotar\"ev has solved the problem \ref{Zolo4} for the simplest case: $E^\pm=\pm[1,1/k]$, $0<k<1$ when $S(x)=sign(x)$.
His solution is given parametrically in terms of elliptic functions and its graph (distorted by a pre- composition with a linear fractional map) is shown in Fig. \ref{ZF} 

\begin{figure}
\centerline{\includegraphics[scale=.7]{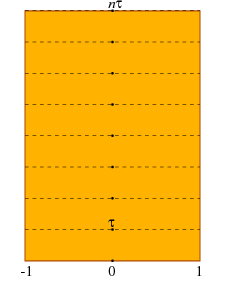}}
\caption{Large rectangle $\Pi_{n\tau}$ composed of $n$ copies of small one $\Pi_\tau$}
\label{Rect}
\end{figure}

To give an explicit representation for this rational function, we consider a rectangle of size $2\times|\tau|$:
$$
\Pi_\tau=\{u\in\mathbb{C}:\quad |Re~u|\le 1, 0<Im~u<|\tau| \},
\qquad \tau\in i\mathbb{R}
$$
which may be conformally mapped to the upper half plane with the normalization
$x_\tau(u):\Pi_\tau, -1,0,1 \longrightarrow \mathbb{H}, -1,0,1$. The latter mapping has a closed appearance
$x_\tau(u)=sn(K(\tau)u|\tau)$ in terms of \emph{elliptic sine} and \emph{complete elliptic integral} 
$K(\tau)=\frac\pi2\theta_3^2(\tau)$, both of modulus $\tau$ \cite{Akh1}. Zolotar\"ev fraction has a parametric representation resembling the definition of a classical Chebysh\"ev polynomial:
$$Z_n(x_{n\tau}(u))=x_\tau(u).$$
Of course,  it takes some effort to prove that $Z_n$ is the rational function of its argument.
The qualitative graph of Zolotar\"ev fraction completely follows from the Fig. \ref{Rect}, for instance its $2n-2$
critical points correspond to the interior intersection points of the vertical boundaries of the large 
rectangle $\Pi_{n\tau}$ and horizontal boundaries of smaller rectangles. Alternation points different from critical points of the fraction 
correspond to four corners of the large rectangle. Zeros/poles of the fraction correspond to $u=l\tau$ with even/odd $l$.

\begin{rmk} Zolotar\"ev fractions share many interesting properties with Chebysh\"ev polynomials as the 
latter are the special limit case of the former  \cite{B12,B17}. For instance, the superposition of suitably chosen Zolotar\"ev
fractions is again a Zolotar\"ev fraction. They also appear as the solutions to many other extremal problems
\cite{Gonchar,Dubini1,Dubini2}.
\end{rmk}

\section{Projective view}
Here we discuss the optimization problem setting which embraces all the formulations we met before in Sect \ref{4Set}. 
We do not treat the infinity point both in the domain of definition and the range of rational function as exceptional. 
Real line extended by a point at infinity becomes a \emph{real projective line} $\mathbb{R}P^1:=\hat{\mathbb{R}}=\mathbb{R}\cup\{\infty\}$
which is a topological circle. We consider two collections of disjoint closed segments on 
the extended real line: $E$ consisting of $m\ge 2$ segments and $F$ of just two segments. The segments of both $E$ and $F$ are of two types:
$E:=E^+\cup E^-$; $F:=F^+\cup F^-$. 

\begin{dfn}
We introduce the class ${\cal R}_n(E,F)$ of real rational functions $R(x)$ of a fixed degree
$\deg R= n$ such that $R(E^+)\subset F^+$ and $R(E^-)\subset F^-$. 
\end{dfn}

\begin{figure}
\begin{picture}(160,22)(-35,0)
\put(5,10){\vector(1,0){80}}
\put(87,9){$\mathbb{R}P^1$}
\put(15,10){\circle*{1}}
\put(10,5){$\partial^-F^-$}
\put(30,10){\circle{1}}
\put(27,5){$\partial^+F^-$}
\put(55,10){\circle*{1}}
\put(50,5){$\partial^-F^+$}
\put(70,10){\circle{1}}
\put(68,5){$\partial^+F^+$}
\put(20,11){$F^-$}
\put(60,11){$F^+$}
\thicklines
\put(15,10){\line(1,0){15}}
\put(55,10){\line(1,0){15}}
\end{picture}
\caption{The ordering of four endpoints $\partial F$ and their colors}
\label{dF}
\end{figure}
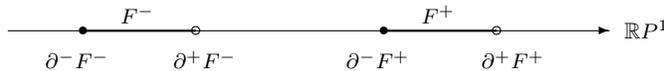

The set of values $F$ modulo \emph{projective (=linear-fractional) transformations} depends on a single value -- 
cross ratio of its endpoints. Suppose the endpoints $\partial F$ are cyclically ordered as follows 
$\partial^-F^-$, $\partial^+F^-$, $\partial^-F^+$, $\partial^+F^+$ -- see Fig. \ref{dF} 
then the cross ratio of four endpoints we define as follows
\begin{dfn}
$$
\kappa(F):=\frac{\partial^+F^+-\partial^+F^-}{\partial^+F^+-\partial^-F^+}:
\frac{\partial^-F^--\partial^+F^-}{\partial^-F^--\partial^-F^+}>1.
$$
\end{dfn}

The classes ${\cal R}_n(E,F)$ (possibly empty) and the value of the cross ratio have several easily checked properties:
\begin{lmm} 1. Monotonicity. 
$$ {\cal R}_n(E,F')\subset {\cal R}_n(E,F)\qquad once ~F'\subset F.$$

2. Projective invariance. For any projective transformations $\alpha,\beta\in PGL_2(\mathbb{R})$,
$$
 {\cal R}_n(\alpha E,\beta F)=\beta\circ {\cal R}_n(E,F)\circ\alpha^{-1}.
$$

3. The value $\kappa(F)$ is decreasing with the growth of its argument: if $F'\subset F$ then $\kappa(F')>\kappa(F)$. 
\end{lmm}

\subsection{Projective problem setting}
Fix degree $n>0$ and set $E\subset\mathbb{R}P^1$, find
$$
\varkappa(n,E):=\inf\{\kappa(F): \qquad {\cal R}_n(E,F)=\emptyset\}. 
$$
The idea behind this optimization is the following: we squeeze the set of values $F$, the functional class $ {\cal R}_n(E,F)$ diminishes
and we have to catch the moment -- quantitatively described by the cross ratio $\kappa(F)$ -- when the class disappears.

\begin{rmk}
1) In problem formulation \ref{Zolo3} the set $F^+=[-\theta,\theta]$ and the set $F^-=[1/\theta,-1/\theta]$; $\kappa(F)=\left(\frac12(\theta+1/\theta)\right)^2$.
In setting \ref{Zolo4} the sets $F^\pm=\pm[1-\mu,1+\mu]$ and $\kappa(F)=\mu^{-2}$. 

2) Notice that the cross ratio depends on the order 
of four participating  endpoints and may take six values interchanged by the elements of the so called unharmonic group.
\end{rmk}

\subsection{Decomposition into subclasses}
Now we decompose each set ${\cal R}_n(E,F)$ into subclasses which were first introduced for the problem setting \ref{minDev} by R.A.-R. Amer 
in his PhD thesis \cite{AS} in 1964. The construction of these subclasses is purely topological: suppose we identify opposite points of a circle $S^1$, we get a double cover of a circle identified with real projective line $\mathbb{R}P^1$ by another circle $S^1$. Now we try to lift the mapping $R(x):\quad \mathbb{R}P^1\to\mathbb{R}P^1$ to the double cover 
of the target space: $\tilde{R}(x):\quad \mathbb{R}P^1\to S^1$. There is a topological obstruction to the existence of $\tilde{R}$: the mapping degree or the winding number of $R(x)$ modulo 2. A simple calculation shows that this value is equal to algebraic degree $\deg R$ $mod~2$. However this lift exists  on any simply connected piece of $\mathbb{R}P^1$. Suppose the segment $E(j)$ is made up of two consecutive segments $E_j,~E_{j+1}$ of the set of bands $E=\cup_{j=1}^m E_j$ and the gap between them. The set $F\subset\mathbb{R}P^1$ lifted to the circle $S^1$ consists of four components cyclically ordered as $F_0^-,F_0^+,F_1^-,F_1^+\subset S^1$. The mapping $R(x):\quad E(j)\to\mathbb{R}P^1$ has two lifts to the covering circle $S^1$ and exactly one of them has values $\tilde{R}(x)\in F_0:=F_0^-\cup F_0^+$ when $x\in E_j$. On the opposite side $E_{j+1}$ of the segment the same function $\tilde{R}(x)$ takes values in the set $F_{\sigma(j)}$ with well defined $\sigma(j)\in\{0,1\}$. Totally, the function $R(x)$ defines an element of $\mathbb{Z}_2$ for any two consecutive segments of the set $E$ with the only constraint
$$
\sum_{j=1}^m \sigma(j) =\deg~R ~~mod ~2. 
$$
which defines the element $\Sigma:=(\sigma_1,\sigma_2,\dots,\sigma_{m-1})\in\mathbb{Z}_2^{m-1}$.
All elements $R(x)\in {\cal R}_n(E,F)$ with the same value of $\Sigma\in\mathbb{Z}_2^{m-1}$ make up a subclass $ {\cal R}_n(E,F,\Sigma)$.
Again, one readily checks the properties of the new classes:

\begin{lmm}
1. Monotonicity:
$$
{\cal R}_n(E,F',\Sigma)\subset {\cal R}_n(E,F,\Sigma) \qquad once ~F'\subset F.
$$
2. Projective invariance
$$
\beta\circ {\cal R}_n(E,F,\Sigma)={\cal R}_n(E,\beta F,\beta\Sigma), \qquad \beta\in PGL_2(\mathbb{R}).
$$
here projective transformation $\beta$ acts on $\Sigma$ component wise: 
$\sigma(j)$ reverses exactly when $\beta$ changes the orientation of projective line
and bands $E_j,E_{j+1}$ are of different $\pm$-type. Otherwise -- iff $\beta\in PSL_2(\mathbb{R})$ or bands $E_j,E_{j+1}$ are both pass- or stopbands -- $\sigma(j)$ is kept intact.
\end{lmm}

\begin{rmk} R.-A.R.Amer \cite{AS} combines classes ${\cal R}_n(E,F,\Sigma)$ and ${\cal R}_n(E,F,\beta\Sigma)$
for  $\beta$ reversing the orientation of projective line and conserving the components $F^\pm$.
This is why he gets twice less number $2^{m-2}$ of classes.
\end{rmk}

\subsection{Extremal problem for classes}
Given  degree $n$, set of bands $E$, and the class $\Sigma$ -- find
\be
\label{minsigma}
\varkappa(n,E,\Sigma):=\inf\{\kappa(F): \qquad {\cal R}_n(E,F,\Sigma)=\emptyset\}. 
\ee

\subsection{Equiripple property}
\begin{dfn}
We say that cyclically ordered (on projective line) points $a_1,a_2,\dots a_s\subset E$  
make up an alternation set for the function $R(x)\in {\cal R}_n(E,F)$
iff $R(x)$ maps any of those points to the boundary $\partial F=\partial^+F\cup\partial^-F$,  and any two consecutive points -- to different sets $\partial^+F$,  $\partial^-F$ colored black and white in Fig. \ref{dF}.
\end{dfn}

\begin{thrm}
If the value $\varkappa(n,E,\Sigma)>1$, then the closure of the extremal class ${\cal R}_n(E,F,\Sigma)$ contains a unique function 
$R(x)$ which is characterized by the property of having at least $2n+2$ alternation points when $R(x)$ is not at the boundary of the class.
\end{thrm}
Proof of this theorem and other statements of the current Section will be given elsewhere.

\section{Problem genesis: signal processing}
There are many parallels between analogue and digital electronics, this is why many
engineering solutions of the past have moved to the new digital era. In particular, 
the same optimization problem for rational functions discussed in Sect. \ref{4Set} 
arises in the synthesis of both analogue and digital electronic devices.

From mathematical viewpoint electronic device is merely a linear operator which transforms 
input signals $x(\cdot)$ to output signals $y(\cdot)$. By signals they mean functions of one continuous or
discrete argument: $x(t)$, $\quad t\in\mathbb{R}$ or $x(k)$, $\quad k\in\mathbb{Z}$. 
For technical simplicity they assume that signals vanish in the `far past'. Another natural assumption 
that a device  processing a delayed signal gives the same but (equally) delayed output which mathematically means that 
operator commutes with the time shifts. As a consequence, the operator consists in a (discrete) convolution of the input signal with 
the certain fixed signal -- the response $h(\cdot)$ to (discrete) delta function input:
$$
y(t)=\int_{\mathbb{R}}h(t')x(t-t')dt'; 
\qquad 
y(k)=\sum_{k'\in\mathbb{Z}}h(k')x(k-k'). 
$$
The causality property means that the output cannot appear before the input and implies that \emph{impulse response} $h(\cdot)$ 
vanishes for negative arguments. Further restrictions on the \emph{impulse response} follow from the physical construction of 
the device.

Analogue device is an electric scheme assembled of elements like resistors, capacitors, (mutual) inductances, etc.  which is governed by 
Kirchhoff laws. Digital device is governed by the recurrence relation:
\be
\label{filterDef}
y(m):=\sum_{j=0}^n p_jx(m-j)+ \sum_{j=1}^n q_jy(m-j),
\qquad m\in\mathbb{Z}.
\ee
To compute the impulse response, we use the Fourier transform of continuous signals and Z-transform of digital ones
(here we do not discuss any convergence):
\be
\label{FZimage}
\hat{x}(\omega):=\int_{\mathbb{R}}x(t)\exp(i\omega t)dt;
\quad\omega\in\mathbb{H}
\qquad
\hat{x}(z):=\sum_{k\in\mathbb{Z}}x(k)z^k,
\quad z\in\mathbb{C}.
\ee
Using the explicit relation \eqref{filterDef} for digital device and its Kirchhoff counterpart for analogue ones we observe that the images of input signals are merely multiplied by \emph{rational functions} $\hat{h}(\cdot)$ of appropriate argument. Since the impulse response is real valued, its image -- also called the \emph{transfer function} -- has the symmetry 
$$
\hat{h}(-\bar{\omega})=\overline{\hat{h}(\omega)},
\qquad 
\hat{h}(\bar{z})=\overline{\hat{h}(z)},
\qquad \omega, z\in\mathbb{C}.
$$
  In practice we can physically observe the absolute value of the transfer function: if we 'switch on' a harmonic signal of a given frequency as the input one, then after certain transition process the output signal will also become harmonic, however with a different amplitude and phase. The magnification of the amplitude as a function of frequency is called the magnitude response function and it is exactly equal to the absolute value of transfer function of the device.

\emph{Multiband filtering} consists in constructing a device which almost keeps the magnitude of a harmonic signal
with the frequency in the passbands and almost eliminates signals with the frequency in the stopbands.
We use the word 'almost' since the square of the magnitude response is a rational function on the real line (for analogue case):
\be
\label{MRf}
|{\hat{h}(\omega)}|^2=\hat{h}(\omega)\hat{h}(-\omega)=R(\omega^2), 
\qquad\omega\in\mathbb{R}. 
\ee
At best we can talk of approximation of an indicator function which is equal $0$ at the stopbands and $1$ at the passbands.
For certain reasons discussed e.g. by W.Cauer, uniform (or Chebyshev) approximation is preferable for this practical problem.
So we immediately arrive at the fourth Zolotarev problem taking the square of frequency as a new variable. For the digital case we 
get a similar problem set on the segments of the unit circle which can be transformed to the problem on a real line.

Note that the reconstruction of the \emph{transfer function} $\hat{h}(\cdot)$ from the \emph{magnitude response} is not unique: we have to 
solve the equation \eqref{MRf} given its right hand side, which has some freedom. This freedom is used to meet another important restriction on the image of impulse response which is prescribed by the causality: $\hat{h}(\cdot)$ can only have poles in the lower half plane $-\mathbb{H}$ of complex variable $\omega$ for the analogue case or strictly outside the unit circle of variable $z$ for the digital case.

Minimal deviation problem in any of the given above settings is just an intermediate step to the following problem of 
great practical importance. \emph{Find minimal degree filter meeting given filter specifications}
like the boundaries of the pass- and stop- bands attenuation at the stopbands and allowable ripple amplitude at the passbands.
The degree of the rational function $\hat{h}$ is directly related to the complexity of device structure, its size,  weight, 
cost of production, energy consumption, cooling, etc... 

\section{Approaches to optimization}
There are three major approaches to the practical solution of optimization problem of multiband 
electrical filter.

\subsection{Remez-type methods} Direct numerical optimization is usually based on Remez-type methods. This is a group
of algorithms specially designed for uniform rational approximation \cite{Remez,Veidinger,Fuchs,8}.
They iteratively build the necessary alternation set for the error function of approximation.
Unfortunately the intrinsic instability of Remez algorithms does not allow to get high degree solutions
and therefore sophisticated filter specifications. For instance, standard double precision accuracy $10^{-15}$
used e.g. in MATLAB does not allow to get solutions of degree $n$ greater than 15-20. We know an  example 
when approximation of degree $n\approx2000$ required mantissa of 150000 decimal signs for 
stability of intermediate computations. Writing just one number of this precision requires 75 standard pages  -- 
this is the volume of a typical PhD thesis.  Another problem of this group of algorithms is the choice of 
initial approximation. The set of suitable starting points may have infinitesimal volume.

\subsection{Composite filters} Practical approach of engineers is to decompose 
complicated problem into many simple ones and solve them one by one. In case of filter synthesis they use 
a  battery of single passband (say, Cauer) filters. This approach is very reliable: it always gives working solutions 
which however are far from being optimal. We get a substantial rise in the order of filter, and 
therefore complexity of its structure and the downgrading of many consumer properties.

\subsection{Ansatz method}  Is based on an explicit analytical formula for the solution generalizing 
formula for Zolotar\"ev fractions. However this formula contains unknown parameters, both continuous and discrete
which have to be evaluated given the input data of the problem. Of all approaches this one is the 
least studied from the algorithmic viewpoint and its usage is restrained by involved mathematical apparatus
\cite{B10}. Nonetheless it copes with very involved filter specifications: narrow transition bands,
large number of working bands, their different proportions, high degree of solution. 

A detailed comparison of three approaches has been made in \cite{BGL}.

\section{Novel analytical approach}
\label{Ansatz}
The idea behind this approach utilizes the following observation: \emph{ Almost all -- with very few exceptions -- critical points of the extremal function have values in some 4-element set $\sf Q$}. Indeed, the  equiripple property claims that a degree $n$ solution has $2n+2$ alternation points, those in the interior of $E$ being critical. Their  values  belong to the set ${\sf Q}:=\pm\theta, \pm1/\theta$ in the settings 1), 2), 3) or $\pm1\pm\mu$ in setting 4) or $\partial F$ for the projective setting. This number is roughly equal to the total
number $2n-2$  of critical points of a degree $n$ rational function. The number $g-1$ of exceptional critical points  is counted as
\be
g-1= \sum\limits_{x: \,
R(x)\not\in{\sf Q}} B(R,x) + \, 
\sum\limits_{x: 
\, R(x)\in{\sf Q}}
\left[
\frac12  B(R,x)
\right],
\label{g-1}
\ee
here the summation is taken over points of the Riemann sphere;
$[\cdot]$ is the integer part of a number and $B(R,x)$ is the branching number of the holomorphic map $R$ at the point $x$. 
The latter value equals zero in all regular points $x$ including simple poles of $R(x)$, or  the multiplicity of the critical point
of $R(x)$ otherwise.

Mentioned above exceptional property of extremal rational functions may be rewritten in a form of a generalized Pell-Abel 
functional equation and eventually gives the desired few-parametric representation of the solution \cite{B10} for the normalized 
${\sf Q}=\{\pm1, \pm1/k(\tau)\}$
\begin{equation}
\label{R} 
R(x)=
\mathop{\rm sn} 
\left(\int_{e}^{x}d\zeta + A(e)\,\biggl|\,\tau\right).
\end{equation}
Here $\mathop{\rm sn}(\cdot|\tau)$ is the \emph{elliptic sine} of the \emph{modulus} $\tau$ related to the value of the 
deviation (depending on the setting it is $\mu, \theta$ or $\kappa(F)$); $d\zeta$ is a \emph{holomorphic differential} on the unknown beforehand  \emph{hyperelliptic curve}
\be
\label{M}
M=M({\bf e})=\{(w,x)\in \mathbb{C}^2:\quad w^2=\prod_{s=1}^{2g+2}(x-e_s)\}, \qquad {\bf e}=\{e_s\}_{s=1}^{2g+2}.
\ee
This curve has branching at the points $e\in\bf e$ where $R(x)$ takes values from the exceptional set $\sf Q$ \emph{with odd multiplicity}. One can show that the genus $g$ of the curve \eqref{M} equals to the above defined number \eqref{g-1} of exceptional critical points plus 1. The arising surface is not arbitrary: it bears a holomorphic differential $d\zeta$ whose periods lie in the rank  2 periods lattice of elliptic sine. The phase shift 
$A(e)$ is some quarter period of $\mathop{\rm sn}(\cdot|\cdot)$.

Algebraic curves of this type are not new to mathematicians:  they are so called \emph{Calogero-Moser curves} and describe the dynamics of points on a torus interacting with the Weierstrass potential $\wp(u)$.

\section{Examples of filter design}
\begin{figure}
\centerline{\includegraphics[scale=.55]{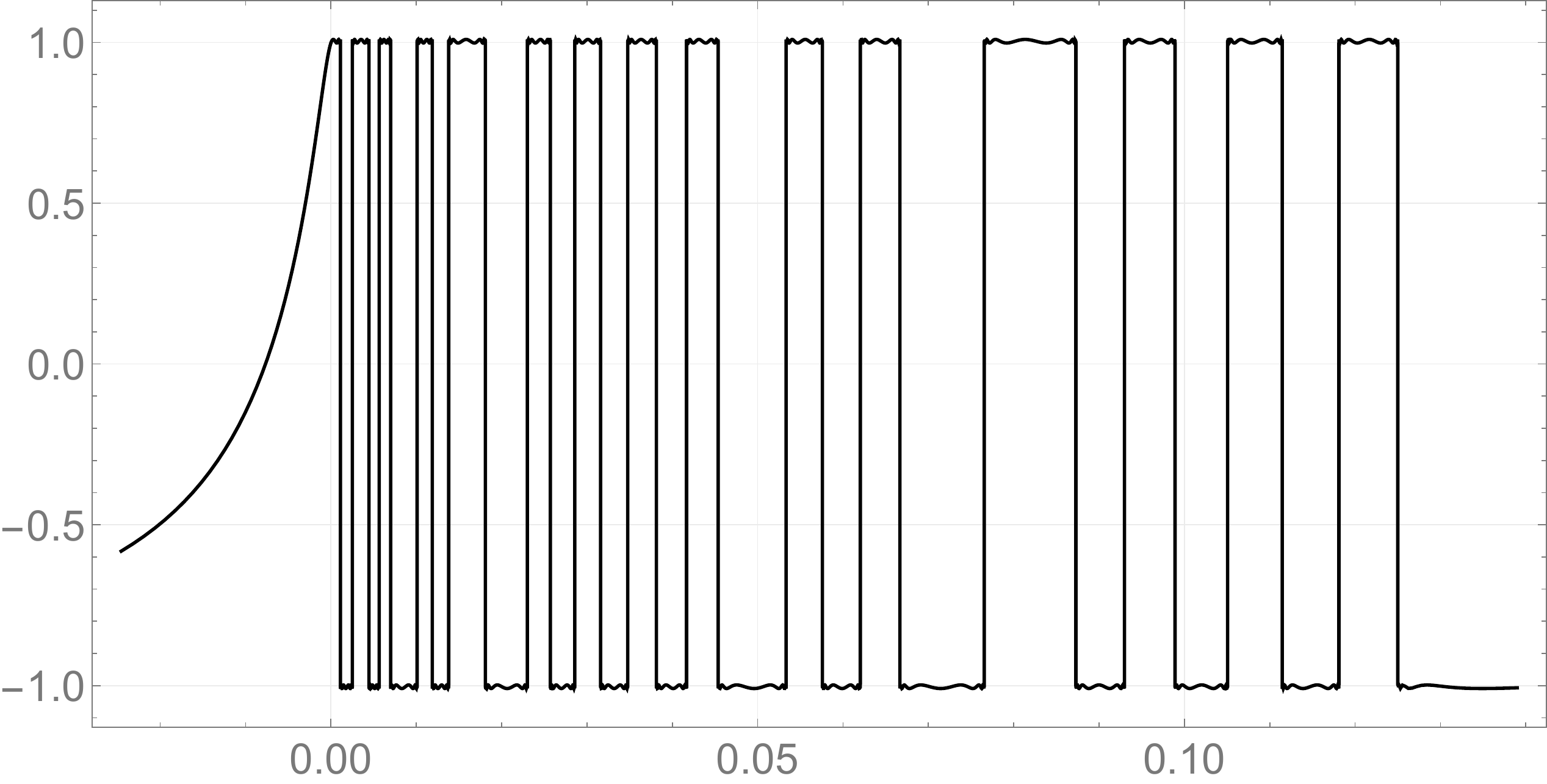} }
\caption{Optimal magnitude response function with 30 work bands.}
\label{30bands}
\end{figure}

\begin{figure}[h!]
\centerline{\includegraphics[scale=.13]{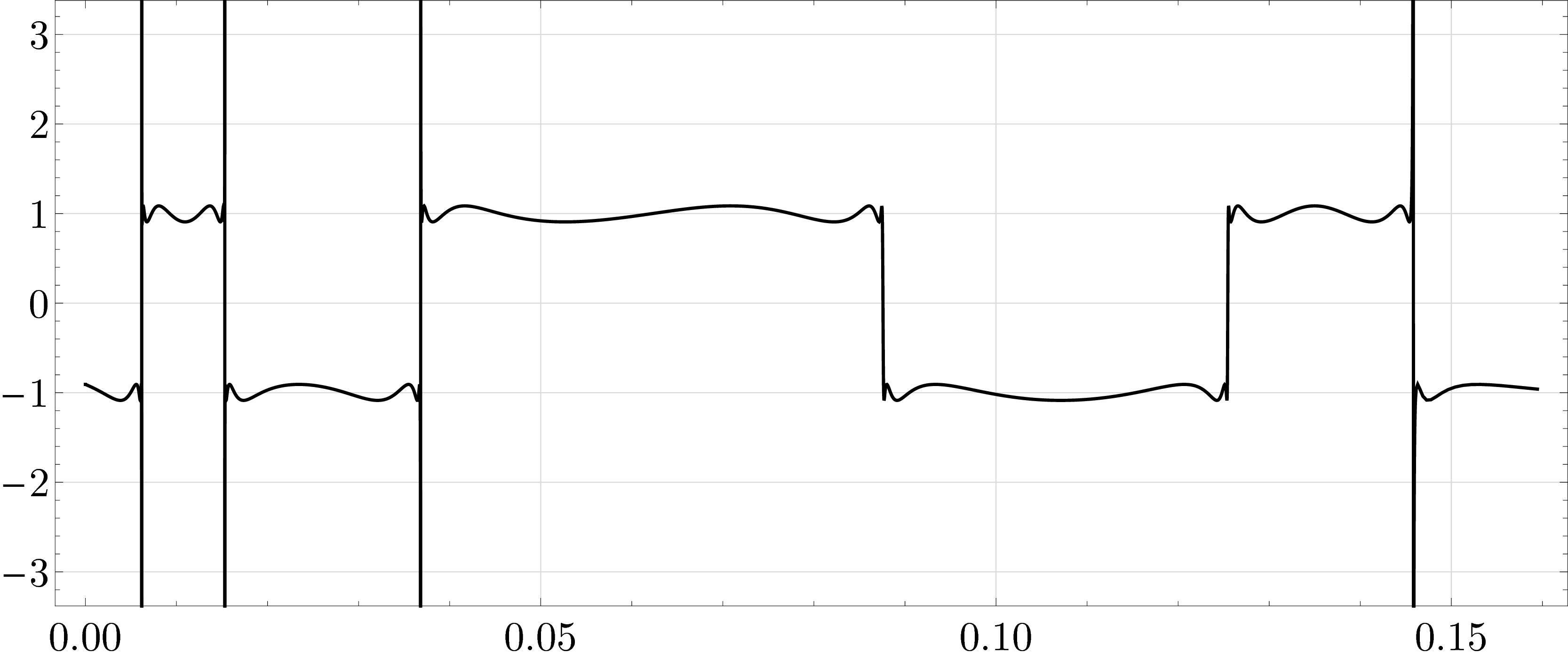} }
\caption{The minimizer for the fourth Zolotar\"ev problem on 7 bands with class $\Sigma$ admitting poles in some transition bands.}
\label{PoleBands} 
\end{figure}

We give several examples of optimal magnitude response functions from different classes, all of them are computed by Sergei Lyamaev.
Fig. \ref{30bands} shows the solution of fourth Zolotar\"ev problem with the set $E$ consisting of 30 bands. 
The solution contains no poles in the transition bands and may be transformed to the transfer function of the multiband filter.
Fig. \ref{PoleBands} shows the solution of the problem with seven working bands. Its class $\Sigma$ admits poles
in the transitions and the function cannot be used for the filter synthesis, which does not exclude other possible applications.

\begin{figure}[h!]
\includegraphics[scale=.07]{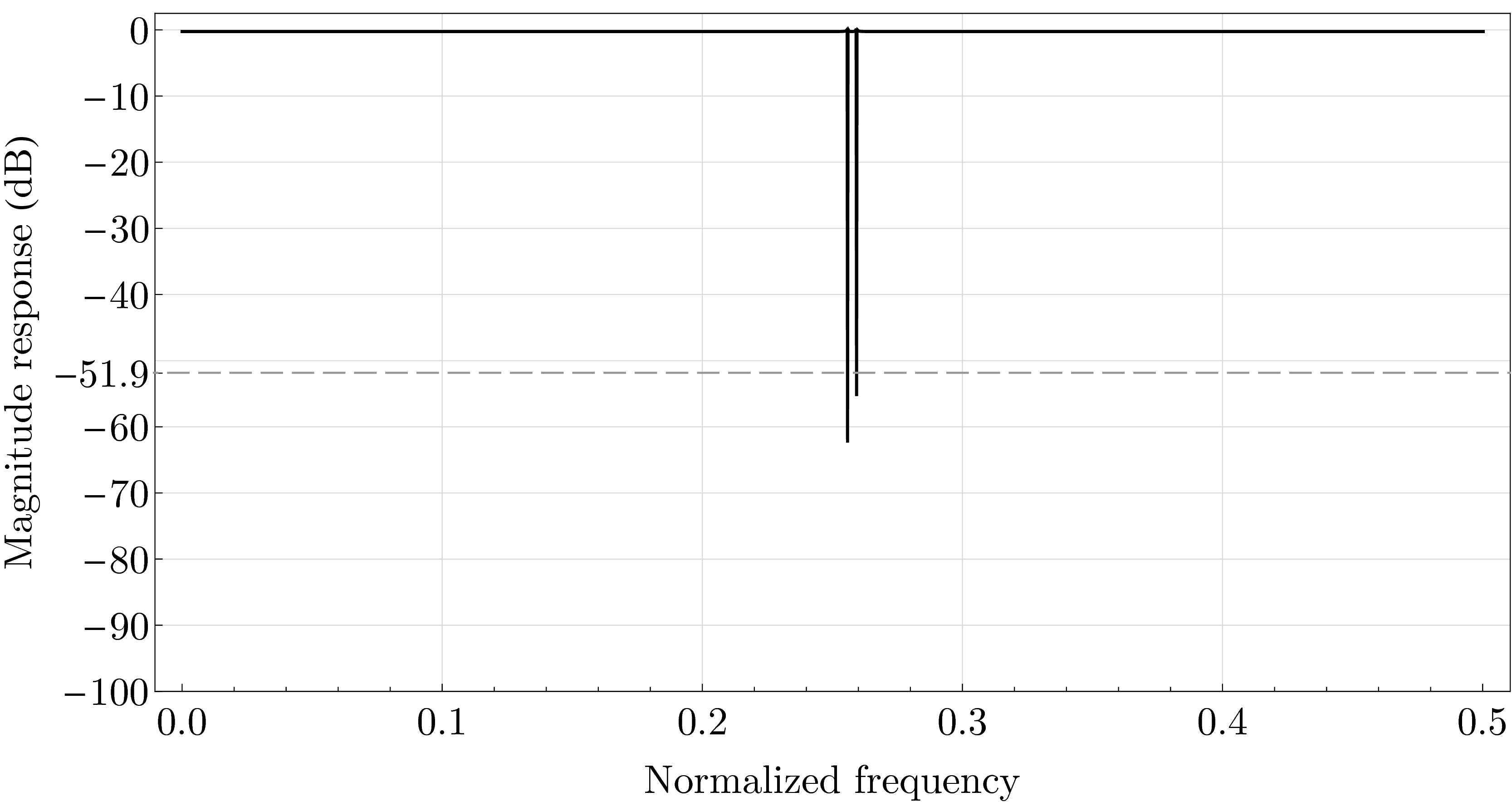}
\hfill
\includegraphics[scale=.07]{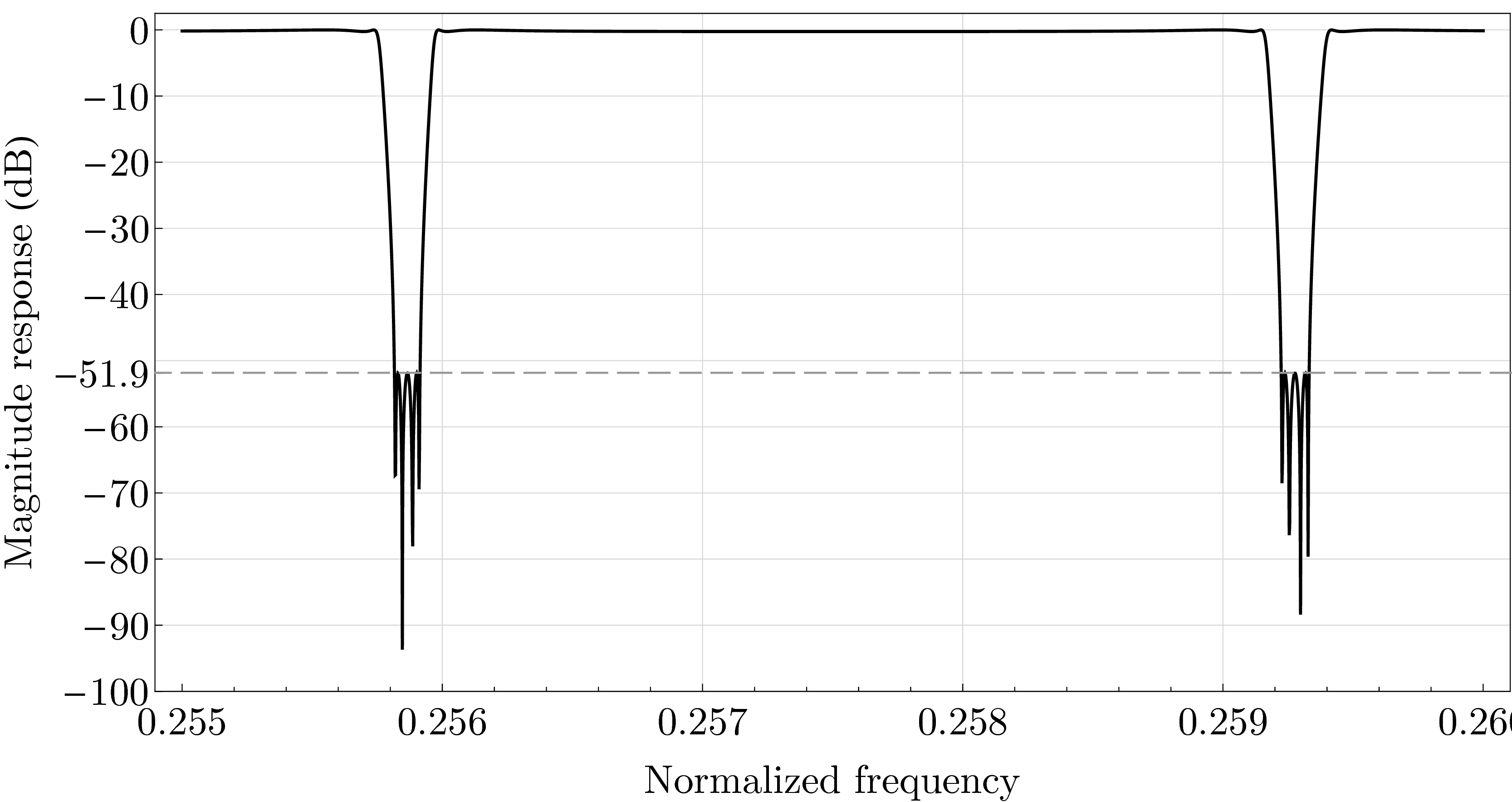}
\caption{Optimal \emph{double notch} filter eliminating noise at two given frequencies (left).
Magnification of the left figure  at two cut off frequencies (right).
Log scale on the vertical axis.}
\label{DNotch} 
\end{figure}

Fig. \ref{DNotch} represents a magnitude response function of the so called double notch filter
which eliminates input signal in the narrow vicinities of two given frequencies.
Shown here optimal filter has degree $n=16$ while same specification composite filter has degree $n=62$.

\end{document}